\documentclass[12pt]{article}
\usepackage{latexsym}
\usepackage{amsmath}
\usepackage{graphicx}
\usepackage{epsfig}
\usepackage{amsfonts}
\usepackage{amssymb}
\begin{document}

\title{Recursive constructions and their maximum likelihood decoding}
\date{}
\author{Ilya Dumer and Kirill Shabunov\thanks{The authors are with the College of
Engineering, University of California, Riverside, CA 92521. This research was
supported by the NSF grant NCR-9703844.}}
\maketitle
\begin{abstract}
We consider recursive decoding techniques for RM codes, their subcodes, and
newly designed codes. For moderate lengths up to 512, we obtain near-optimum
decoding with feasible complexity.
\end{abstract}

\section{Introduction}

\thispagestyle{empty}In this paper, we consider decoding algorithms that can
achieve good performance and low complexity on moderate blocklengths. Our goal
is to fill the void left by the best algorithms, such as optimum maximum
likelihood (ML) decoding, which has unfeasible complexity even on relatively
short blocks, and iterative decoding, which becomes very efficient beginning
with the lengths of tens of thousands. More specifically, we wish to achieve
near-optimum performance on the lengths ranging from 128 to 512, where neither
of these two algorithms can yet combine good performance with low complexity.\smallskip

To achieve this goal, we will use \textit{recursive} \textit{techniques. }One
particular class of codes generated by (multilevel) recursion is Reed-Muller
(RM) codes and their subcodes. Also, RM codes are only slightly inferior to
the best codes on moderate lengths. We will see below that recursive decoding
substantially outperforms other (nonexponenential) algorithms known for RM
codes. Our basic recursive procedure will split the $RM$ code $(r,m)$ of
length $n$ into the two constituent RM codes $(m$-$1,r$-$1)$ and $(m$-$1,r)$
of length $n/2$. Decoding is then relegated further to the shorter codes until
we reach basic codes with feasible ML decoding. In all intermediate steps, we
only recalculate the reliabilities of the newly defined symbols.

To improve decoding performance, we will also generalize recursive design. In
particular, we use subcodes of RM codes and their modifications. We also use
relatively short lists of \ code candidates in the intermediate steps of \ the
recursion. As a result, we closely approach ML decoding performance on the
blocklengths up to 512.

\section{Reed-Muller codes}

\noindent We use notation $\left\{
\genfrac{}{}{0pt}{}{m}{r}%
\right\}  $ for RM codes $(n,k)$ of length $n=2^{m},$ dimension $k=\sum
_{i=0}^{r}{\binom{m}{i}}$ and distance $d=2^{m-r}$. \ RM codes found numerous
applications thanks to fast decoding procedures. First, \textit{majority
algorithm }\cite{ree} enables feasible bounded-distance decoding and can even
correct \cite{kri} most error patterns of weight up to $(d\ln d)/4$ on long
codes of fixed rate $R.$ \smallskip

Other efficient decoding schemes are based on recursive technique of
\cite{lit} and \cite{kab}. These algorithms enable bounded distance decoding
with the lowest complexity order of $n\min(r,m-r)$ known for RM codes.
\thinspace\thinspace Simulation results \cite{bos} show that recursive
algorithms increase decoding domain of bounded distance decoding.\thinspace
\ Subsequently, these algorithms were slightly refined in \cite{dum4}. It was
shown that (similar to majority decoding) recursive algorithms of \cite{lit}
and \cite{kab} correct most error patterns up to the weight $(d\ln d)/4$ when
used on long codes of fixed rate $R$.\smallskip

For long low-rate RM codes of fixed order $r$, both majority decoding and
recursive schemes correct most error patterns of Hamming weight up to
$n(1-\varepsilon_{r}^{\text{maj}})/2,$ where the residual term has vanishing
order
\begin{equation}
\varepsilon_{r}^{\text{maj}}\sim(m/d)^{1/2^{r+1}} \label{eps-maj}%
\end{equation}
as $m\rightarrow\infty$. Note that (\ref{eps-maj}) gives a
\textit{threshold-type}\textbf{\ }capacity that approaches the upper limit of
$n/2.$ However, degree of convergence is relatively slow even for codes
$\left\{
\genfrac{}{}{0pt}{}{m}{2}%
\right\}  .$ Much better results are obtained for ML decoding. For long codes
$\left\{
\genfrac{}{}{0pt}{}{m}{r}%
\right\}  $ of fixed order $r,$ it is proven in \cite{sid} that ML decoding
further reduces the residual term $\varepsilon_{r}^{\text{maj}}$ to the order
of
\begin{equation}
\varepsilon_{r}^{\text{\textsc{ML}}}\precsim(\ln4)\sqrt{m^{r}/n}.
\label{transit}%
\end{equation}

\section{Recursive structure}

In essence, all recursive techniques known for RM codes are based on the
\textit{Plotkin construction}. Here the original RM code $\left\{
\genfrac{}{}{0pt}{}{m}{r}%
\right\}  $ is represented in the form $(u\mathbf{,}u\mathbf{+}v),$ by taking
any subblock\textbf{\ }$u$ from RM $\left\{
\genfrac{}{}{0pt}{}{m-1}{r}%
\right\}  $ and any $v$ from RM $\left\{
\genfrac{}{}{0pt}{}{m-1}{r-1}%
\right\}  .$ These two subcodes have length $2^{m-1}.$ By continuing this
process, we again obtain the shorter RM codes of length $2^{m-2}$ and so on.
Finally, we arrive at the end nodes that are repetition codes $\left\{
\genfrac{}{}{0pt}{}{j}{0}%
\right\}  $ and full spaces $\left\{
\genfrac{}{}{0pt}{}{j}{j}%
\right\}  .$This is schematically shown in Fig. 1 for RM codes of length 32.
In Fig. 2, we consider incomplete decomposition terminated at the biorthogonal
codes $\left\{
\genfrac{}{}{0pt}{}{j}{1}%
\right\}  $ and single-parity check codes $\left\{
\genfrac{}{}{0pt}{}{j}{j-1}%
\right\}  $.\smallskip

Now let $I%
\genfrac{}{}{0pt}{}{m}{r}%
$ denote a block of $\ $information bits \ that encodes a vector $(u,u+v).$ It
is also important that our recursion splits $I%
\genfrac{}{}{0pt}{}{m}{r}%
$ into two information subblocks $I%
\genfrac{}{}{0pt}{}{m-1}{r}%
$ and $I%
\genfrac{}{}{0pt}{}{m-1}{r-1}%
$ that encode vectors $u$ and $v\mathbf{,}$ respectively. Correspondingly,
code dimensions satisfy the recursion $|I%
\genfrac{}{}{0pt}{}{m}{r}%
|=|I%
\genfrac{}{}{0pt}{}{m-1}{r}%
|+|I%
\genfrac{}{}{0pt}{}{m-1}{r-1}%
|$. In this way, the shorter information subbloks can be split again until we
arrive at the end nodes$.$ Thus, any specific codeword can be encoded from the
(multiple) information strings assigned to the end nodes $\left\{
\genfrac{}{}{0pt}{}{j}{0}%
\right\}  $ or $\left\{
\genfrac{}{}{0pt}{}{j}{j}%
\right\}  $.\ Following \cite{kab}, it can be proven that recursive encoding
of code $\left\{
\genfrac{}{}{0pt}{}{m}{r}%
\right\}  $ has complexity
\begin{equation}
\psi_{r}^{m}\leq n\min(r,m-r)+1. \label{encoding}%
\end{equation}
This observation comes from two facts. First, the end nodes $\left\{
\genfrac{}{}{0pt}{}{j}{0}%
\right\}  $ and $\left\{
\genfrac{}{}{0pt}{}{j}{j}%
\right\}  $ satisfy the bound (\ref{encoding}). Second, consider the two
constituent codes $\left\{
\genfrac{}{}{0pt}{}{m-1}{r-1}%
\right\}  $ and $\left\{
\genfrac{}{}{0pt}{}{m-1}{r}%
\right\}  .$ \ Then $(u,u+v)$ construction gives complexity $\psi%
\genfrac{}{}{0pt}{}{m-1}{r-1}%
+\psi%
\genfrac{}{}{0pt}{}{m-1}{r}%
+\frac{n}{2}$ for the code $\left\{
\genfrac{}{}{0pt}{}{m}{r}%
\right\}  .$ \ Using this recursion, one can show that $\psi%
\genfrac{}{}{0pt}{}{m}{r}%
$ also satisfies (\ref{encoding}) if constituent codes do.\vspace{0.1in}

\ \ \ \ \ \ \ \ \ \ \ \ \ \ \ \ \ \ \ \ \ \ \ \ \ \ \ \ \ \ \ {\small 0,0}%
\ \ \ \ \ \ \ \ \ \ \ \ \ \ \ \ \ \ \ \ \ \ \ \ \ \ \ \ \ \ \ \ \ \ \ \ \ \ \ \ \ \ \ \ \ \ \ \ \ \ \ \ \ {\small 2,1}%
\ 

\ \ \ \ \ \ \ \ \ \ \ \ \ \ \ \ \ \ \ \ \ \ \ \ \ \ \ \ $\nearrow$
$\ \nwarrow$%
\ \ \ \ \ \ \ \ \ \ \ \ \ \ \ \ \ \ \ \ \ \ \ \ \ \ \ \ \ \ \ \ \ \ \ \ \ \ \ \ \ \ \ \ \ \ \ \ \ $\nearrow
$ $\ \nwarrow$

\ \ \ \ \ \ \ \ \ \ \ \ \ \ \ \ \ \ \ \ \ \ \ \ \ {\small 1,0}{\large \ \ \ }%
\ {\large \ }\ {\small 1,1 }%
\ \ \ \ \ \ \ \ \ \ \ \ \ \ \ \ \ \ \ \ \ \ \ \ \ \ \ \ \ \ \ \ \ \ \ \ \ \ \ \ \ \ {\small 3,1}%
{\large \ \ \ }\ {\large \ }\ \ {\small 3,2}

$\ \ \ \ \ \ \ \ \ \ \ \ \ \ \ \ \ \ \ \ \ \ \nearrow$ \ $\nwarrow
${\large \ \ \ }\ $\nearrow$ $\ \nwarrow$
$\ \ \ \ \ \ \ \ \ \ \ \ \ \ \ \ \ \ \ \ \ \ \ \ \ \ \ \ \ \ \ \ \ \ \ \ \nearrow
$ \ $\nwarrow${\large \ \ \ }\ $\nearrow$ $\ \nwarrow$

\ \ \ \ \ \ \ \ \ \ \ \ \ \ \ \ \ \ {\small 2,0}{\large \ \ \ }\ {\large \ \ }%
\ {\small 2,1}{\large \ \ \ }\ {\large \ \ }\ \ {\small 2,2
\ \ \ \ \ \ \ \ \ \ \ \ \ \ \ \ \ \ \ \ \ \ \ \ \ \ \ \ \ \ \ 4,1{\large \ \ \ }%
\ {\large \ \ \ \ }4,2{\large \ \ \ }\ {\large \ \ }\ 4,3\ \ \ \ \ \ \ \ }

\ \ \ \ \ \ \ \ \ \ \ \ \ \ \ \ $\nearrow$ $\ \nwarrow${\large \ \ \ }%
\ $\nearrow$ $\ \nwarrow${\large \ \ \ }\ $\nearrow$ $\ \nwarrow
$\ \ \ \ \ \ \ \ \ \ \ \ \ \ \ \ \ \ \ \ \ \ \ \ $\nearrow$ $\ \nwarrow
${\large \ \ \ }\ $\nearrow$ $\ \nwarrow${\large \ \ \ }\ $\nearrow$
$\ \nwarrow$ \ 

\ \ \ \ \ \ \ \ \ \ \ \ \ {\small 3,0}{\large \ \ \ \ \ \ }{\small 3,1}%
{\large \ \ \ \ \ \ \ }\ {\small 3,2}{\large \ \ \ }\ {\large \ \ }%
\ {\small 3,3 \ \ \ \ \ \ \ \ \ \ \ \ \ \ \ \ \ \ \ \ \ 5,1{\large \ \ \ }%
\ {\large \ \ }5,2{\large \ \ \ \ \ \ }\ \ 5,3{\large \ \ \ \ \ \ }5,4\ }

\ \ \ \ \ \ \ \ \ $\nearrow$ $\ \nwarrow${\large \ \ \ }\ $\nearrow$
$\ \nwarrow${\large \ \ \ }\ $\nearrow$\ $\ \nwarrow${\large \ \ \ }%
\ $\nearrow$ $\ \nwarrow$\ \ \ \ \ 

\ \ \ \ \ \ {\small 4,0}{\large \ \ \ \ \ \ }\ {\small 4,1}{\large \ \ \ }%
\ {\large \ }\ \ \ {\small 4,2}{\large \ \ \ }\ {\large \ \ }\ \ {\small 4,3}%
{\large \ \ \ }\ {\large \ \ }\ {\small 4,4 \ \ \ \ \ \ }

\ \ \ $\nearrow$ $\ \nwarrow${\large \ \ \ }\ $\nearrow$ $\ \nwarrow
${\large \ \ \ }\ $\nearrow$ $\ \nwarrow${\large \ \ \ }\ $\nearrow$
$\ \nwarrow${\large \ \ \ }\ $\nearrow$ $\ \nwarrow$ \ \ \ 

{\small 5,0}{\large \ \ \ }\ {\large \ \ \ }{\small 5,1}{\large \ \ \ \ \ }%
{\small \ \ \ \ 5,2}{\large \ \ \ \ \ }\ {\large \ }{\small 5,3}%
{\large \ \ \ }\ {\large \ \ \ }\ {\small 5,4}{\large \ \ \ \ }\ {\large \ }%
\ {\small 5,5 \ \ \ \ \ }

\smallskip

Fig. 1$:$ Full decomposition\qquad\qquad\qquad\qquad\qquad\ \ \ Fig. 2$:$
Partial decomposition

\section{\ New decoding techniques}

\noindent Our algorithm also uses the $(u\mathbf{,}u\mathbf{+}v)$ construction
and relegates decoding to the two constituent RM codes. Decoder receives a
block $(\widetilde{u},\widetilde{u\mathbf{+}v})$ that consists of two halves
$\widetilde{u}$ and $\widetilde{u\mathbf{+}v}$ corrupted by noise. We first
try to find the better protected codeword $v$ from $\left\{
\genfrac{}{}{0pt}{}{m-1}{r-1}%
\right\}  \mathbf{.}$ Then we proceed with the block \ $u$ from the code
$\left\{
\genfrac{}{}{0pt}{}{m-1}{r}%
\right\}  $.\ In a more general scheme, we repeat this recursion, by
decomposing subblocks $v$ and $u$ further. On all intermediate steps, we only
recalculate the probabilities of the newly defined symbols. Finally, we
perform soft decision ML decoding once we reach the end nodes. The most
important difference from the previous work \cite{dum4} is that in each step
we keep $L$ most probable candidates obtained prior to this step. This
difference is discussed in Section 5. \ In this section, we first assume that
our decoding is terminated on the biorthogonal codes depicted in Fig. 2$.$

\textbf{Step 1}. \ To find a subblock $v$ in hard-decision decoding, one would
use its corrupted version $\widetilde{v}=\widetilde{u}+\widetilde{u+v}.$ Using
more general approach, we find the posterior probabilities of \ the received
symbols. On the left half $\widetilde{u},$ \textbf{\ }each symbol $u_{i}$ has
posterior probability%
\[
p_{i}^{\prime}\overset{\text{def}}{=}\Pr\{u_{i}=0\,\,|\widetilde{u_{i}}\,\}.
\]
Similarly, we use the right half $\widetilde{u+v}$ to find the posterior
probability of any symbol $u_{i}+v_{i}:$%
\[
p_{i}^{\prime\prime}\overset{\text{def}}{=}\Pr\{u_{i}\mathbf{+}v_{i}%
=0\,\,|\,\widetilde{u_{i}\mathbf{+}v_{i}}\}.
\]
Given the probabilities $p_{i}^{\prime}$ and $p_{i}^{\prime\prime}$ of the
symbols $u_{i}$ and $u_{i}\mathbf{+}v_{i},$ we then find the posterior
probability $p(v_{i})$ of their binary sum $v_{i}.$ \ Here we use the formula
of total probability and find
\begin{equation}
p(v_{i})\overset{\text{def}}{=}\Pr\{v_{i}=0\,\,|\,\,\,\widetilde{u_{i}%
}\,,\widetilde{u_{i}\mathbf{+}v_{i}}\}=p_{i}^{\prime}p_{i}^{\prime\prime
}+(1-p_{i}^{\prime})(1-p_{i}^{\prime\prime}). \label{1}%
\end{equation}
Here we use the fact that the two original symbols $u_{i}$ and $u_{i}%
\mathbf{+}v_{i}$ are independent$.$ Also, both symbols are independently
corrupted by Gaussian noise. \ Now we can use any soft-decision decoding that
uses probabilities $p(v_{i})$ to find the most probable vector $v$ from the
$\left\{
\genfrac{}{}{0pt}{}{m-1}{r-1}%
\right\}  $-code. This completes Step 1 of our algorithm. Vector $v$ is then
passed to Step 2.

\textbf{Step 2}. \ Now we use both vectors $\ \widetilde{u+v}$ and $v$ to
estimate each symbol $u_{i}$ on the right half. \ \textit{Assuming that} $v$
\textit{is correct}, we find \ that each symbol $u_{i}$ has posterior
probability
\[
p_{i}^{\wedge}\overset{\text{def}}{=}\Pr\{u_{i}=0\,\,|\,\,\widetilde
{u_{i}\mathbf{+}v_{i}},\,v_{i}\}=\left\{
\begin{array}
[c]{ll}%
p_{i}^{\prime\prime}, & \text{if}\;v_{i}=0,\\
1-p_{i}^{\prime\prime}, & \text{if}\;v_{i}=1.
\end{array}
\right.
\]
Now we have the two posterior probabilities $p_{i}^{\prime}$ and
$p_{i}^{\wedge}$ of symbols $u_{i}$ obtained on both corrupted halves. By
using the Bayes' rule, we find the combined estimate
\begin{equation}
p(u_{i})\overset{\text{def}}{=}\Pr\{u_{i}=0\,|\,\,p_{i}^{\prime}%
,\,p_{i}^{\wedge}\}=\frac{p_{i}^{\prime}p_{i}^{\wedge}}{p_{i}^{\prime}%
p_{i}^{\wedge}+(1-p_{i}^{\prime})(1-p_{i}^{\wedge})}. \label{3}%
\end{equation}
Finally, we perform soft decision decoding and find a subblock\ $u\mathbf{\in
}\left\{
\genfrac{}{}{0pt}{}{m-1}{r}%
\right\}  .$ \medskip

Thus, procedure $\left\{
\genfrac{}{}{0pt}{}{m}{r}%
\right\}  $ has a recursive structure that calls procedures $\left\{
\genfrac{}{}{0pt}{}{m-1}{r-1}%
\right\}  $ and $\left\{
\genfrac{}{}{0pt}{}{m-1}{r}%
\right\}  ,$ and so on. By recalculating probabilities (\ref{1}) and
(\ref{3}), we finally arrive at the \textit{biorthogonal} Reed-Muller codes
$\left\{
\genfrac{}{}{0pt}{}{j}{1}%
\right\}  $ on our way to the left, or full codes $\left\{
\genfrac{}{}{0pt}{}{j}{j}%
\right\}  $ on the way to the right. \textit{Maximum likelihood} decoding is
executed on the end nodes. Each decoding retrieves a new subset of information
symbols associated with the current end node.\ In both cases, maximum
likelihood decoding has complexity order at most $n\log_{2}n$ \cite{lit3}.
\ Simple analysis also shows that recalculating all posterior probabilities in
(\ref{1}) and (\ref{3}) has complexity at most $5n.$ Therefore our decoding
complexity $\Psi%
\genfrac{}{}{0pt}{}{m}{r}%
$ satisfies the recursion
\[
\Psi_{r}^{m}\leq\Psi_{r-1}^{m-1}+\Psi_{r}^{m-1}+5n.
\]
This brings the overall complexity to the order of $5n\log_{2}n$ real
operations. A slightly more efficient version gives complexity $(5n\log_{2}n)/2.$

\section{Analysis and improvements}

Given the code $\left\{
\genfrac{}{}{0pt}{}{m}{r}%
\right\}  ,$ we first decode code $\left\{
\genfrac{}{}{0pt}{}{m-r+1}{1}%
\right\}  $ followed by codes $\left\{
\genfrac{}{}{0pt}{}{m-r}{1}%
\right\}  ,$ $\left\{
\genfrac{}{}{0pt}{}{m-r-1}{1}%
\right\}  ,$ and so on. With the exception of the leftmost and the rightmost
nodes, the procedure enters each node multiple times, by taking all the paths
leading to this node. It turns out that the output bit error rate (BER)
significantly varies on different nodes and even on different paths leading to
the same node. Therefore our first problem is to define the most error-prone
paths. We start our analysis with two examples.\medskip\ 

\textbf{Example 1. } For simplicity, assume that the all-zero codeword from
the code $\left\{
\genfrac{}{}{0pt}{}{m}{r}%
\right\}  $ is transmitted over the binary channel with crossover probability
$p=0.9$. \ Then we use formula (\ref{1}) with $p_{i}^{\prime}=p_{i}%
^{\prime\prime}=0.9$ to find the probability $p(v_{i})$ of correct symbol
$v_{i}=0$ in the block $v\in\left\{
\genfrac{}{}{0pt}{}{m-1}{r-1}%
\right\}  .$ From (\ref{1}) we see that $p(v_{i})=0.82$. Subsequently, this
probability $p(v_{i})$ rapidly converges to 0.5 in \ a few more steps. On the
positive side, we note that each step gives us a better protected code that
has twice the relative distance of the former one. In particular, \ the
leftmost node $\left\{
\genfrac{}{}{0pt}{}{m-r+1}{1}%
\right\}  $ has length $2d$ and distance $d.$ Its ML decoding gives
asymptotically vanishing BER if the residual term $\varepsilon_{1}%
^{\text{\textsc{ML}}}$ still exceeds $(\ln4)\sqrt{m/2d}$, according to
(\ref{transit}).\medskip

\textbf{Example 2. } Suppose that our original code $\left\{
\genfrac{}{}{0pt}{}{m}{r}%
\right\}  $ from the previous example has already received the correct
subblock $v$ from the code $\left\{
\genfrac{}{}{0pt}{}{m-1}{r-1}%
\right\}  .$ Now we need to find the remaining subblock $u$ from the code
$\left\{
\genfrac{}{}{0pt}{}{m-1}{r}%
\right\}  .$ Given correct $v,$ we can use (\ref{3}) with $p_{i}^{\prime
}=p_{i}^{\wedge}=0.9.$ Then we find that symbol $u_{i}$ is correct with
probability $p(u_{i})\approx0.99$. \ Now we see that the probability
$p(u_{i})$ rapidly increases as we move to the right. Note, however, that each
new code has half the relative distance of its parent code. In other words, we
subsequently improve the channel while entering the new codes with weaker
correcting capabilities. Finally, the last code $\left\{
\genfrac{}{}{0pt}{}{r}{r}%
\right\}  $ has no error protection and gives the output BER equal to its
input error probability. \medskip

\textbf{Asymptotic analysis. }For AWGN channels, we assume that the all-zero
codeword is transmitted as a sequence of $\ +1$s. Then we receive $n$
independent random variables $\,$(RV) $\widetilde{u_{i}}\,$\ and
$\widetilde{u_{i}\mathbf{+}v_{i}}$ with normal distribution $\mathcal{N}%
(1,\sigma^{2}).$ Accordingly, it can be readily seen that the posterior
probabilities $p_{i}^{\prime}$ (and $p_{i}^{\prime\prime})$ become independent
RV with non-Gaussian distribution (here $\tanh(x)$ is hyperbolic tangent):
\begin{equation}
p_{i}^{\prime}=(1+\varepsilon_{i}^{\shortmid})/2,\qquad\text{where
}\varepsilon_{i}^{\shortmid}=\tanh(2\widetilde{u_{i}}/\sigma^{2}).
\label{prob1}%
\end{equation}
In the next step, we obtain the RV $p(v_{i})$ and \ $p(u_{i}).$ Their
distributions can also be written (see \cite{dum4}) in the form (\ref{prob1}),
where the residual terms are:
\begin{equation}
\varepsilon_{i}(v_{i})=\varepsilon_{i}^{\shortmid}\varepsilon_{i}%
^{\shortparallel},\qquad\varepsilon_{i}(u_{i})=\tanh(2\widetilde{u_{i}}%
/\sigma^{2}+(-1)^{v_{i}}2(\widetilde{u_{i}\mathbf{+}v_{i}})/\sigma^{2}).
\label{recalc}%
\end{equation}
Similar to Example 1, it can be shown that the first RV $\varepsilon_{i}%
(v_{i})$ has a smaller expectation $\overline{\varepsilon_{i}(v_{i})}$
relative to the original estimate $\varepsilon_{i}^{\shortmid}$. By contrast,
the second RV $\varepsilon_{i}(u_{i})$ has a greater expected value. Here the
analysis is similar to Example 2. \ We also use that the newly defined RV
$\varepsilon_{i}(v_{i})$ and $\varepsilon_{i}(u_{i})$ are all independent for
each new step. Now consider asymptotic case of high noise power $\sigma^{2}%
\gg1$. (Note that this case is relevant to long RM codes with $m\rightarrow
\infty$ and fixed order $r.)$\ \ Then we use asymptotic approximations in
formulas (\ref{1}) and (\ref{3}) and arrive at the following conclusions.\medskip

$\bullet$ We prove that moving to the left from $\left\{
\genfrac{}{}{0pt}{}{m}{r}%
\right\}  $ to $\left\{
\genfrac{}{}{0pt}{}{m-1}{r-1}%
\right\}  $ and further is equivalent to squaring our noise power (bringing it
to $\sigma^{4}$, then $\sigma^{8},$ and so on), while keeping the signal
energy equal to 1. By contrast, the original noise power $\sigma^{2}$ is cut
by half when the algorithm moves to the right (bringing it to $\sigma^{2}/2$,
$\sigma^{2}/4$, and so on).

$\bullet$ We prove that the left-hand movement makes our subcodes much more
vulnerable. In this case, doubling the relative code distance $d/n$ does not
compensate for a stronger noise$.$ In particular, the highest (worst) BER
$\mathcal{P}_{1}$ is obtained on the leftmost node $\left\{
\genfrac{}{}{0pt}{}{m-r+1}{1}%
\right\}  $ that is decoded first. The second worst BER $\mathcal{P}_{2}$ is
obtained on the next decoded node $\left\{
\genfrac{}{}{0pt}{}{m-r}{1}%
\right\}  ,$ and so on. \ Using conventional notation $Q(x)=\int_{x}^{\infty
}e^{-u^{2}/2}du/\sqrt{2\pi},$ we prove that for $m\rightarrow\infty:$%
\begin{equation}
\mathcal{P}_{1}\sim Q(2^{(m-r)/2}\sigma^{-2^{r-1}}),\qquad\mathcal{P}_{2}\sim
Q(2^{(m-r+1)/2}\sigma^{-2^{r-1}}). \label{prob-out}%
\end{equation}
Now we see that even two adjacent nodes give very different results, where
$\mathcal{P}_{2}\sim\mathcal{P}_{1}^{2}$ for small $\mathcal{P}_{1}.$ By
contrast, moving to the right does not increase the output BER relative to the
parent code. In this case, the lowest BER is obtained on the rightmost node
$\left\{
\genfrac{}{}{0pt}{}{r}{r}%
\right\}  .$\medskip

\noindent\textbf{Asymptotic comparison. }\ For long RM codes new recursive
decoding increasingly outperforms both the majority algorithm and the former
recursive techniques of \cite{lit}, \cite{bos} as the block length grows. In
particular, these algorithms give BER \ $\mathcal{P}\sim Q(2^{(m-r)/2}%
\sigma^{-2^{r}}).$ Further, it can be shown that for long RM codes of fixed
rate $R$, the above decoding corrects most error patterns of weight up to
$(d\ln d)/2$ thus:

$\bullet$ increasing $\ln d$ times the capacity of bounded-distance decoding$;$

$\bullet$ doubling the capacity $(d\ln d)/4$ of the former recursive technique.\medskip

\noindent\textbf{Improvements. }An important conclusion resulting from the
above analysis is to set the leftmost information bits as zeros. In this way,
we arrive at the subcodes of the original code $\left\{
\genfrac{}{}{0pt}{}{m}{r}%
\right\}  $ that are obtained by eliminating only a few least protected
information bits. In particular, even eliminating the first $m-r+2$
information bits that form the leftmost code $\left\{
\genfrac{}{}{0pt}{}{m-r+1}{1}%
\right\}  ,$ immediately can reduce the output BER from $\mathcal{P}_{1}$ to
its square $\mathcal{P}_{2}$ for sufficiently long codes.\smallskip

Decoding performance can be further improved by using \textit{list decoding}.
To simplify the analysis, we now consider the repetition codes $\left\{
\genfrac{}{}{0pt}{}{j}{0}%
\right\}  $. In particular, we start with the leftmost \ code $\left\{
\genfrac{}{}{0pt}{}{m-r}{0}%
\right\}  $ and take \textit{both} codewords $v=0$ and $\overline{v}=1$ of
length $2^{m-r}.$ Correspondingly, we keep both posterior probabilities
instead of choosing the more probable codeword. \ This step gives the two
initial edges of \ a tree. Each edge is associated with a \ cost function
equal to the log of the (corresponding) posterior probability. \smallskip

Then we decode the next code $\left\{
\genfrac{}{}{0pt}{}{m-r-1}{0}%
\right\}  .$ Note that the former codewords $v$ and $\overline{v}$ give
different probability distributions on this node. Given $v$ and $\overline
{v},$ our new decoding is performed 2 times, separately for $v$ and
$\overline{v}.$ \ The result is a full tree of depth 2, that has 4 new edges
along with their cost functions. The next step includes 4 decodings of the
code $\left\{
\genfrac{}{}{0pt}{}{m-r-2}{0}%
\right\}  $ performed \ on each path of the tree. By continuing this process,
we arrive at the codes $\left\{
\genfrac{}{}{0pt}{}{1}{0}%
\right\}  $ and $\left\{
\genfrac{}{}{0pt}{}{1}{1}%
\right\}  $. We also keep accumulating the posterior probabilities of our
paths. It can be seen that the resulting $2^{m-r+2}$ paths give full
biorthogonal code $\left\{
\genfrac{}{}{0pt}{}{m-r+1}{1}%
\right\}  .$ \ Choosing the best path at this point becomes equivalent to the
original termination at the biorthogonal codes.

To improve our decoding, we keep all $L$ paths instead of selecting the best
paths.\smallskip\ In a more general scheme, the threshold $L$ can be greater
or smaller than $2^{m-r+2}.$ In any case, we start at the repetition codes and
keep doubling\footnote{We can also increase the number of paths, say, to $4L$
or $8L$ on the nodes RM $(j,j)$.} the number of paths until $2L$ paths are
formed. After $2L$ paths are constructed, we choose $L$ paths with $L$ maximum
cost functions. \ In the end, the most probable path (that is, the path with
the maximum cost function) is chosen among $L$ paths survived at the rightmost node.\smallskip

Both the simulation results and calculations show that continuous regeneration
of $L$ best candidates improves our original algorithm that selected the best
path at each node. In other words, keeping the longer paths allows us to
better separate the transmitted vector from the remaining candidates. As a
result, we substantially reduce the overall BER even when compared to the
expurgated subcodes. Note, however, that our list decoding increases
complexity $L$ times, to the order of $Ln\log_{2}n.$ To refine this scheme
further, recall that the channel quality constantly improves as we move from
the left to the right. Therefore, we can choose the variable threshold $L$
that becomes smaller as our decoding progresses to the rightmost nodes. \ In
this way, we can substantially reduce our list-decoding complexity even when
$L$ originally exceeds $n.$\smallskip

\noindent\textbf{Simulation results.} Our results are described below in
Figures 4 to 9. These figures also reflect the drastic improvements obtained
when both techniques - using the subcodes and short decoding lists - were
combined. The curves with $L=1$ show the performance of \ the refined version
of \ the former recursive techniques from \cite{lit}, \cite{kab}, and
\cite{bos}. For codes of length 256 and 512, the results are now improved by
3.5 to 5 dB at BER $10^{-4}.$\smallskip

While using the maximum lists depicted on each figure, simulation also showed
that in most cases of incorrect decoding, the erroneous result is more
probable than the transmitted vector. This fact shows that our block ER
$\ $(BL ER) is very close to that of ML decoding. In turn, this gives a new
(experimental) bound on the BL ER of ML decoding. Also, our results
substantially surpass other codes with similar parameters \ (see the current
\ ``world records'' on http://www331.jpl.nasa.gov/\texttt{)}. In Fig. 9, we
summarize the results on block ER of ML decoding for RM codes $\left\{
\genfrac{}{}{0pt}{}{8}{2}%
\right\}  $ to $\left\{
\genfrac{}{}{0pt}{}{8}{6}%
\right\}  $ of length 256.\smallskip

It is also interesting that subcodes usually achieve near-ML decoding using
much smaller lists relative to the original RM codes. In particular, a subcode
(256,78) approaches near-ML decoding using \textit{only 32 intermediate
paths}. Note that even one of the most efficient algorithms developed in
\cite{hart2} uses about $10^{5}$ paths for BCH codes of length 256. On the
other hand, our simulation results show that codes of length 512 approach ML
decoding using much bigger lists than codes of length 256. \ To extend the
results for longer codes, we use slightly different constructions described in
the next section.

\section{More general recursive constructions}

\textbf{Multiple splitting of RM codes. }Here we wish to change the original
Plotkin representation. Namely, one can apply more sophisticated partitions
that directly split RM codes in $4,8,$ or more codes of shorter lengths. For
example, by applying Plotkin construction two times, we can split the original
block into four quarters $(u,u+w_{1},u+w_{2},u+w_{1}+w_{2}+v).$ Here $u$ is
taken from the least protected code $\left\{
\genfrac{}{}{0pt}{}{m-2}{r}%
\right\}  ,$ vectors $w_{1}$ and $w_{2}$ belong to the medium-protected code
$\left\{
\genfrac{}{}{0pt}{}{m-2}{r-1}%
\right\}  ,$ while $v$ is taken from the best protected code $\left\{
\genfrac{}{}{0pt}{}{m-2}{r-2}%
\right\}  .$ Simulation performed for this construction did not improve the
results presented in Figures 4 to 9.\smallskip

Slightly better results were obtained for low SNR, when these four codes were
combined in a different way as $(u,u+w_{1},u+w_{1}+w_{2},u+w_{1}+w_{2}+v).$
\ It can be proven that for low rates the latter construction gives asymptotic
improvement to our original Plotkin representation. \ This conclusion stems
from the following facts. As before, the code $v$ is decoded first and is most
vulnerable in recursive decoding. Note also that $v$ is obtained directly in
one step, by adding the third quarter $u+w_{1}+w_{2}$ and the forth quarter
$u+w_{1}+w_{2}+v$ of our construction. Asymptotically, such a step squares the
noise power, as described above. On the other hand, we reduce the length four
times in each step. Accordingly, the new recursive construction reaches the
leftmost nodes $\left\{
\genfrac{}{}{0pt}{}{j}{0}%
\right\}  $ in $r/2$ steps instead of $r$ steps used before. As a result, we
can replace the former term $\sigma^{-2^{r-1}}$ in (\ref{prob-out}) by the
greater term $\sigma^{-2^{r/2}}.$\smallskip

Despite substantial asymptotic improvements, simulation showed that these
improvements start accumulating only on the lengths of 2048 and above.\vspace{0.1in}

\noindent\textbf{Alternating recursions. }Suppose that we use the Plotkin
construction $(u,u+v)$ in Fig. 1$,$ but change our original code $v$ from
$\left\{
\genfrac{}{}{0pt}{}{m-1}{r-1}%
\right\}  $ to $\left\{
\genfrac{}{}{0pt}{}{m-1}{r-2}%
\right\}  .$ In other words, we move one more step to the left relative to the
Plotkin construction as shown in Fig. 3. As a result, the new code $v$ has a
better error protection. This alteration also doubles the distance of code $v$
and gives unequal error protection for the original code. On the other hand,
we also reduce the overall code rate and the SNR per channel symbol (given the
same SNR per information bit). This lower rate can eliminate the advantages of
\ the better protection. To increase code rate in $v$, we then add extra
symbols in the next splitting step. For example, we split $v$ into codes
$\left\{
\genfrac{}{}{0pt}{}{m-2}{r-3}%
\right\}  $ and $\left\{
\genfrac{}{}{0pt}{}{m-2}{r-1}%
\right\}  ,$ by taking one more step to the right as presented in Fig. 3.\newpage

\qquad\ \ \ \ \ {\small m-2,r-3\ \ \ \ m-2,r-1\ \ \ \ m-2,r}

{\small \qquad\ }$\ \ \ \ \ \ \ \ \ \ \ \nwarrow\nearrow^{\longrightarrow}$
$\ \ \ \nwarrow$$\nearrow$

{\small \qquad\ \ \ \ \ \ \ \ \ \ \ m-1,r-2\ \ \ \ \ \ \ \ m-1,r\ }

{\small \qquad\ \ \ \ \ \ \ \ \ \ }$\ ${\small \ \ \ \ }$\ ^{\longleftarrow
}\nwarrow\nearrow$%
$\ \ \ \ \ \ \ \ \ \ \ \ \ \ \ \ \ \ \ \ \ \ \ \ \ \ \ \ \ \ \ \ \ \ \ \ \ \ \ \ \ \ \ \ \ \nwarrow
\nearrow^{\longrightarrow}$$\nwarrow\nearrow\vspace{-0.05cm}$

{\small \qquad\ \ \ \ \ \ \ \ \ \ \ \ \ \ \ \ \ \ \ \ \ m,r\ \ \ \ \ \qquad
\qquad\qquad\qquad\qquad\qquad\ \ \ \ \ \ \ \ \ \ \ \ }$^{\longleftarrow
}\nwarrow\nearrow${\small \vspace{0.1in}}

Figure 3a$:$ Alternating decompositions\qquad\qquad Figure 3b$:$ Underlying
structure\vspace{0.1in}

\noindent Note that in general alternating construction, we can no longer use
RM codes. These only form the first ``building blocks'', such as $\left\{
\genfrac{}{}{0pt}{}{j}{0}%
\right\}  $ and $\left\{
\genfrac{}{}{0pt}{}{j}{j}%
\right\}  $. By contrast, various nodes $\left\{
\genfrac{}{}{0pt}{}{j}{i}%
\right\}  $ only label the edges/paths that correspond to our new codes. The
first simulation results obtained in this direction used an $(u,u+v)$%
-combination of \ $\left\{
\genfrac{}{}{0pt}{}{8}{1}%
\right\}  $ and $\left\{
\genfrac{}{}{0pt}{}{8}{3}%
\right\}  $ codes instead of the original $\left\{
\genfrac{}{}{0pt}{}{9}{3}%
\right\}  $ code. Even this simple combination improved the original code at
low SNR. \ More sophisticated constructions similar to the one of Fig. 3a also
outperform RM codes. However, the alternating constructions that we considered
to date have not yet improved the performance of subcodes presented in Figures
4 to 9.

\begin{figure}[ptb]
\begin{center}
\includegraphics[height=2.9in, width=4.0145in]{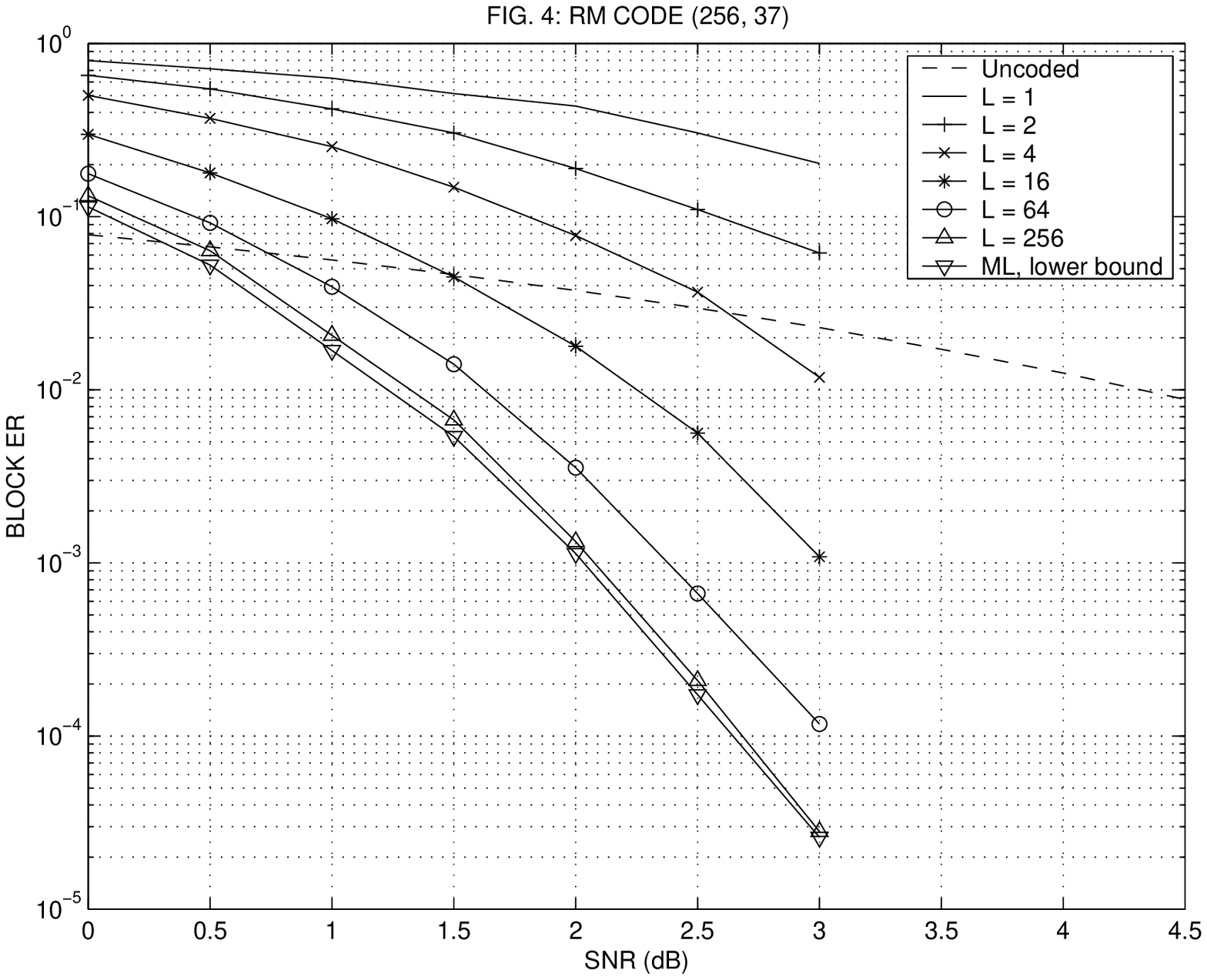}
\end{center}
\end{figure}\begin{figure}[ptbptb]
\begin{center}
\includegraphics[height=3in, width=4.0145in]{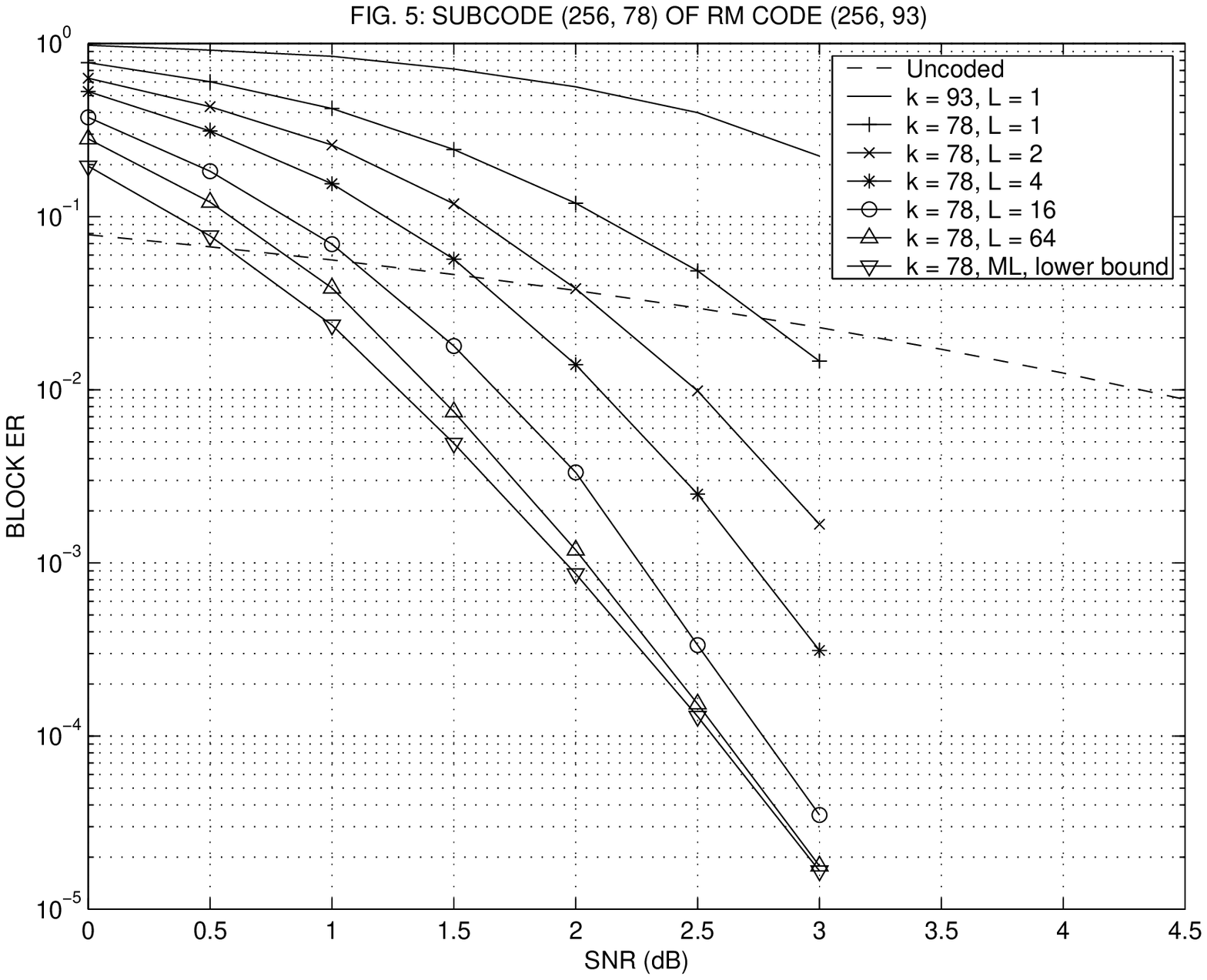}
\end{center}
\end{figure}\begin{figure}[ptbptbptb]
\begin{center}
\includegraphics[height=3in, width=4.0145in]{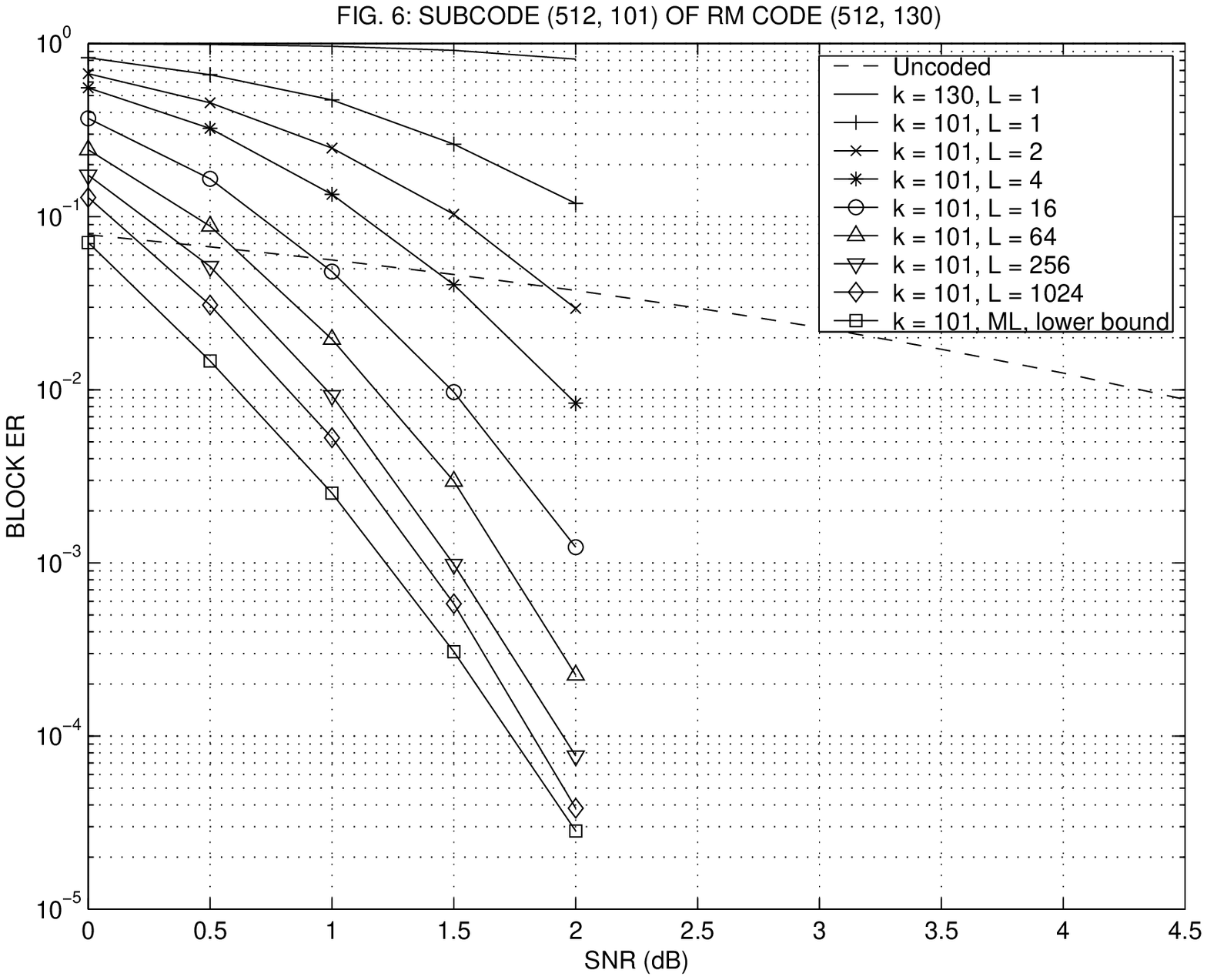}
\end{center}
\end{figure}\begin{figure}[ptbptbptbptb]
\begin{center}
\includegraphics[height=3in, width=4.0145in]{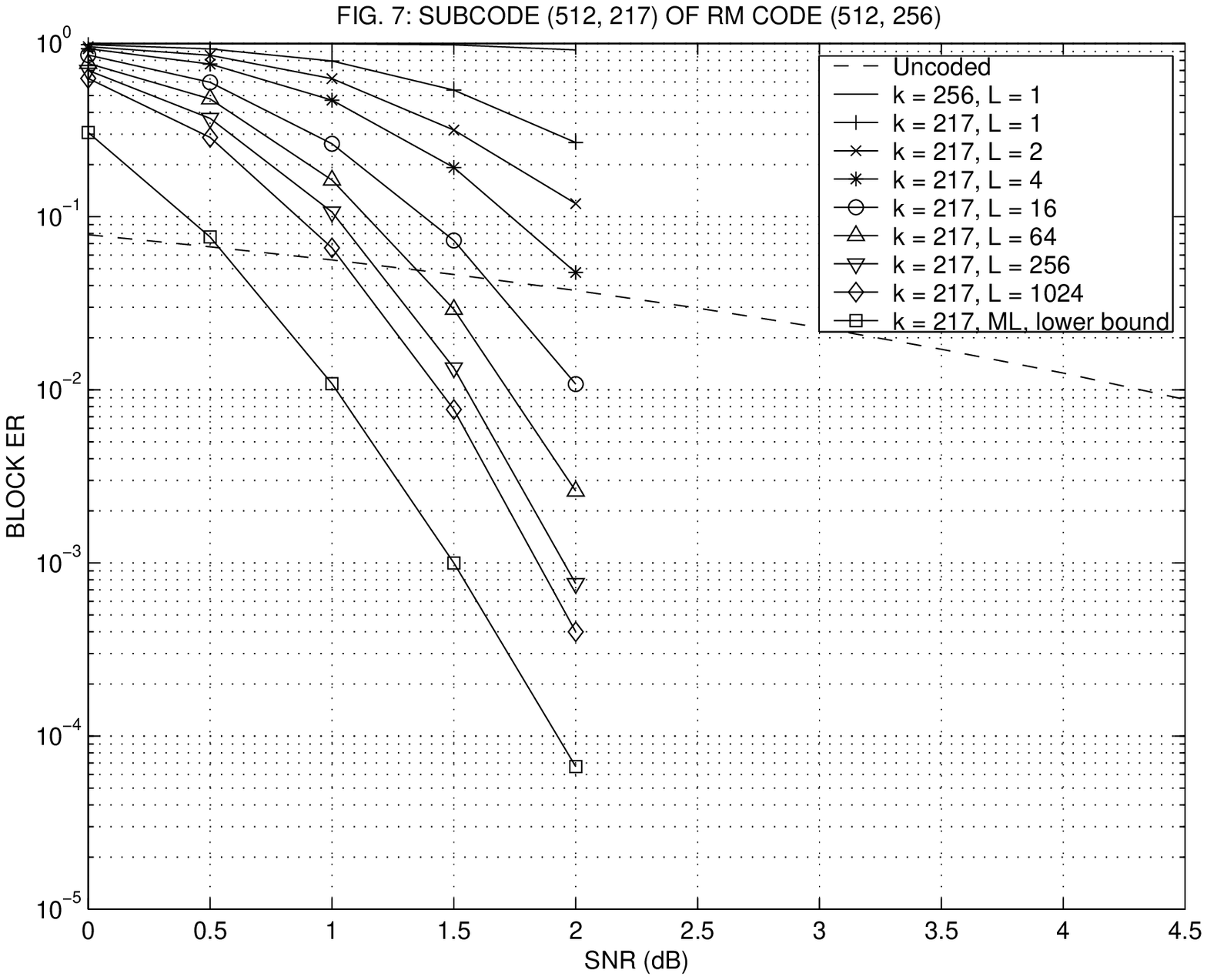}
\end{center}
\end{figure}\begin{figure}[ptbptbptbptbptb]
\begin{center}
\includegraphics[height=2.9in, width=4.0145in]{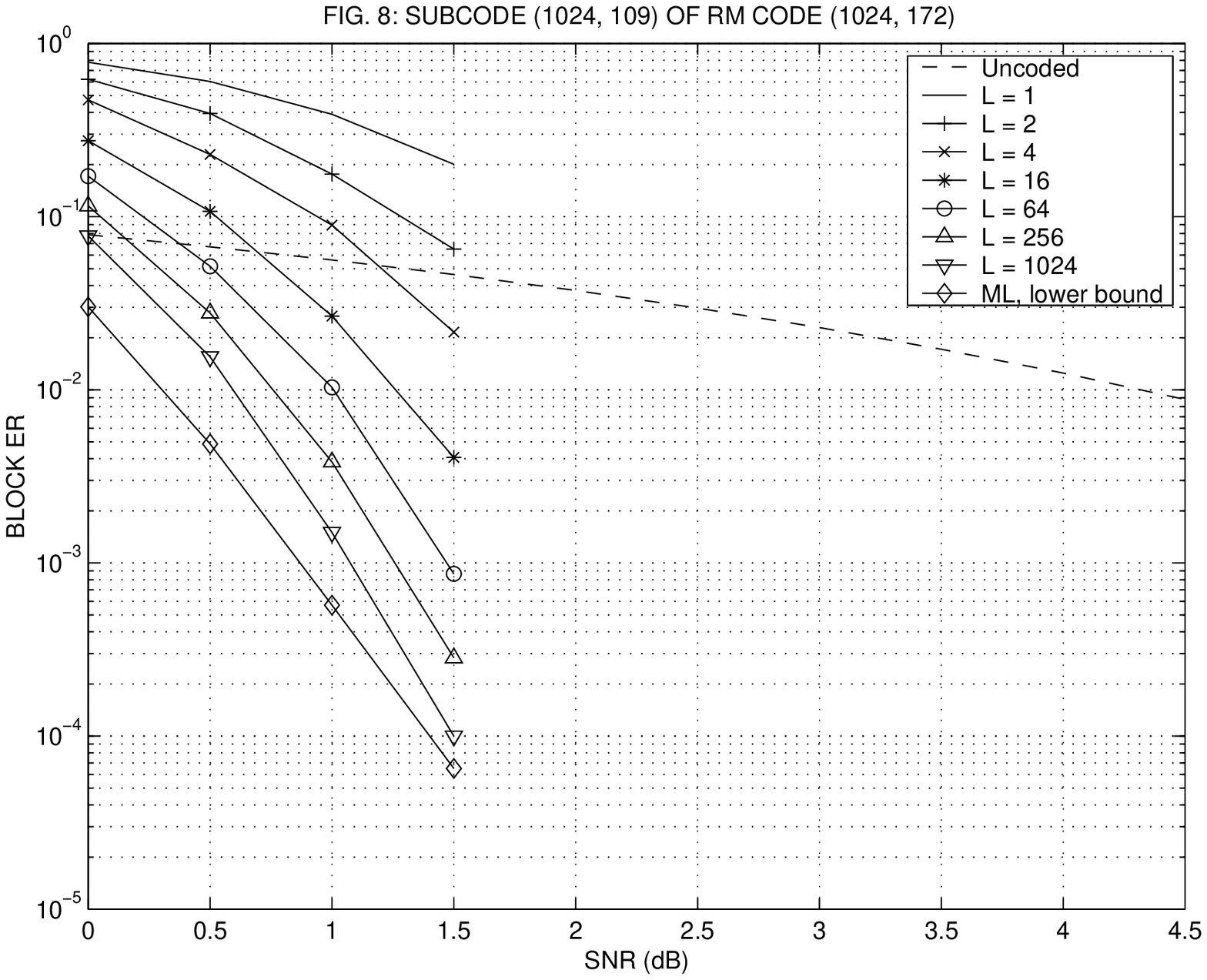}
\end{center}
\end{figure}\begin{figure}[ptbptbptbptbptbptb]
\begin{center}
\includegraphics[height=3in, width=4.0145in]{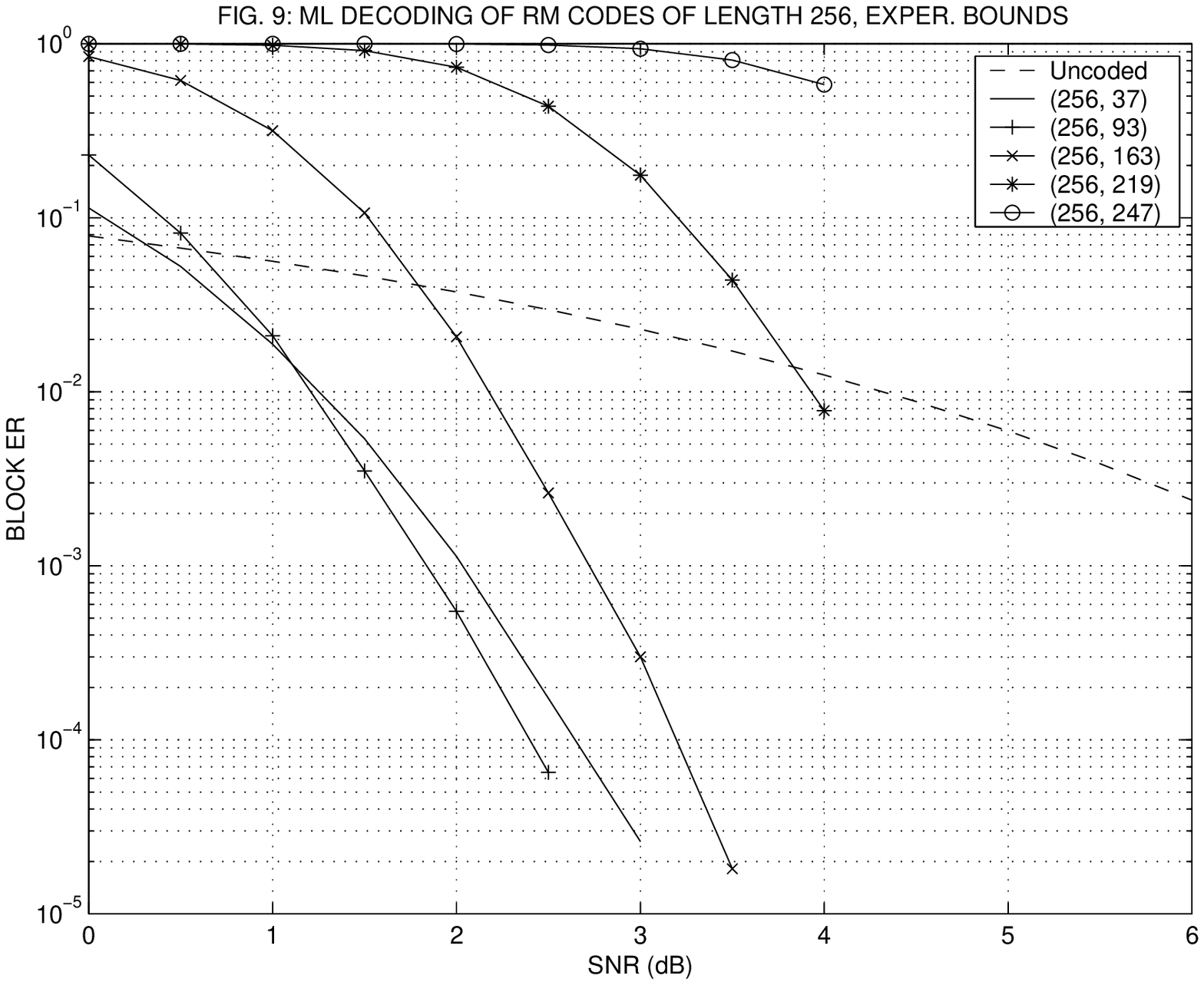}
\end{center}
\end{figure}
\end{document}